\documentclass[journal,twocolumn]{IEEEtran}

\ifCLASSINFOpdf
\else
\fi
\usepackage{graphicx}

\usepackage[english]{babel}
\usepackage[latin1,utf8]{inputenc}
\usepackage[T1]{fontenc}
\usepackage{graphicx}
\usepackage[cmex10]{amsmath}
\usepackage{amssymb}
\usepackage{bm}
\usepackage{cite}
\usepackage{xcolor}
\usepackage{mathtools}
\newcommand{\defeq}{\vcentcolon=}
\usepackage{algorithm}
\usepackage[noend]{algpseudocode}
\usepackage{array}

\makeatletter
\def\BState{\State\hskip-\ALG@thistlm}
\makeatother

\begin{document}

\title{Interference Alignment for Downlink Cellular Networks: Joint Scheduling and Precoding}

\author{Yasser Fadlallah$^{1,2}$,  
       Paul Ferrand$^3$,
	   Leonardo S. Cardoso$^{1,2}$,
       and~Jean-Marie Gorce$^{1,2}$\\
$^{1}$ University of Lyon, INRIA, France\\
$^{2}$ INSA-Lyon, CITI-INRIA, F-69621, Villeurbanne, France\\
$^3$ Huawei Technologies Co. Ltd., French Research Center, F-92659, Boulogne-Billancourt, France\\
$\{$yfadlallah@gmail.com$\}$}

\pagestyle{empty}
\pagenumbering{gobble}
\maketitle

\begin{abstract}
Interference Alignment (IA) is technique that, in a large sense, makes use of the increasing signal dimensions available in the system through MIMO and OFDM technologies in order to globally reduce the interference suffered by users in a network. In this paper, we address the problem of downlink cellular networks, the so-called interfering broadcast channels, where mobile users at cell edges may suffer from high interference and thus, poor performance. Starting from the downlink IA scheme proposed by Suh \textit{et al.}, a new approach is proposed where each user feeds back multiple selected received signal directions with high signal-to-interference gain. A exhaustive search based scheduler selects a subset of users to be served simultaneously, balancing between sum-rate performance and fairness, but becomes untractable in dense network scenarios where many users send simultaneous requests. 
Therefore, we develop a sub-optimal scheduler that greatly decreases the complexity while preserving a near-optimal data rate gain. More interestingly, our simulations show that the IA scheme becomes valuable only in correlated channels, whereas the matched filtering based scheme performs the best in the uncorrelated scenarios.
\end{abstract}

\IEEEpeerreviewmaketitle

\section{Introduction}
Increasing the data rate in wireless networks usually comes through a larger bandwidth, or a more efficient use of the available bandwidth. In cellular applications, the performance is interference-limited in most cases, meaning that increasing the transmission power does not substantially improve the network capacity. On the other hand, separating users in the time or frequency domain leads to very inefficient usage of the available resources. The key challenge is thus to balance interference avoidance and spectrum reuse to reach an optimal trade-off between spectral and energy efficiency~\cite{Tsilimantos2015}.

This challenge has been addressed in the past, for instance using frequency/code planning in 2G/3G networks or with cooperative multiple point antennas in 4G \cite{CoMP}. Dynamic interference management then became a strategic option to not only improve the spectral efficiency but also to achieve greater overall energy efficiency. The IA concept has been proposed first by \cite{MA} and extended by \cite{CJ}, it gave an interesting approach for exploiting interference in a K-users interference channel (IC) situation
The theoretical achievements of IA have been largely discussed, e.g. in \cite{MA,AG,CJ,TG}. One of the key results is that, for a number of theoretical setups, IA can transform interference limited networks into interference-free networks, regardless of the number of users $K$ \cite{CJ}. 

In the downlink cellular network, the IA extension has been addressed in several research works, e.g. \cite{DA,RT,WG,TL,SuhTse}. For example, \cite{RT} proposed to form clusters of base stations (BSs) assuming a global knowledge of channel state information (CSI). As results, only users in the center of clusters benefit from slight data rate improvement
Other results showed questionable performance gains and several limitations \cite{DA}. Among the major issues is that IA sacrifices half of the space dimensions in order to avoid interference for every user in the cluster, whereas some of them may not suffer strong interference. 

More efficiently, Suh \textit{et al.} in \cite{SuhTse} proposed a dynamic precoding scheme that attempts to balance the performance gain of matched filtering for the best users with IA for cell-edge users. By using a fixed rank-deficient precoding step at the transmitter, each BS can ensure that users always see a subspace in their effective channel where interference is reduced or eliminated. The idea is that each mobile measures and feeds back its own free subspace to its main BS, which jointly schedules the UEs so as to maximize the overall capacity. This scheme is particularly attractive in the context of dense cellular networks because with IA and thanks to channel properties, the freed subspace perceived by interfered mobiles is different for each of them. And hence, it becomes more probable to find users for which optimal transmission spaces are quasi-orthogonal as the size of the network increases. Therefore, by performing simultaneously users scheduling and precoding, the BS highly increases the chance to maximize the system capacity.  

The aforementioned IA scheme only assumes to feedback the interference-free direction. However, the additional signal directions may also present a low interference level due to fading on the interfering channel. Therefore, we propose that in addition to the interference-free directions, the users feed back alternative directions, which increases the number of candidate directions at the BS and makes the scheduler more complex. In this paper, a heuristic scheduling process is proposed to decrease the algorithmic complexity while preserving near-optimal performance. Moreover, we focus on a numerical evaluation of the IA performance compared to the matched filtering (MF). Surprisingly, our simulations show that the IA technique gains interest when the crossed channels are correlated while the MF technique performs the best in the uncorrelated case scaling up linearly with the signal-to-interference and noise ratio (SINR).

Notations: boldface upper case letters and boldface lower case letters denote matrices and vectors, respectively. The superscripts $(.)^{\dagger}$ stands for the transpose conjugate matrices, respectively. $\bm I_n$ denotes the identity matrix of dimension $n$. The $\ell_2$ norm is denoted as $||.||$.

\section{System model}
A downlink cellular network with $B$ BS and $N_{m}$ user equipments (UEs) in the $m^{th}$ BS is considered. The $l^{th}$ UE in the $m^{th}$ BS receives $S_{lm}$ streams. All BSs and UEs are equipped with $M$ antennas. The transmission scheme is based on a MIMO-OFDM with $K$ available sub-carriers yielding a total $M_K = M \times K$ dimensions for the transmitted signal, in both frequency and spatial domain. A complex matrix $\bm P$ of dimension $M_K \times (M_K-N_f)$ is used at each BS ($N_f\geq 0$). The data symbols at a given BS are carried out, each by a precoding vector $\bm v_{lm}^j$ ($l, m, j$ denote the user, the BS, and the stream indices, respectively), and then further precoded using $\bm P$. Independent flat fading channels are assumed for all subcarriers. The transmitted signal of the $m^{th}$ BS is given by
\begin{equation}\label{eq1}
\bm{x}_m=\sum_{l=1}^{N_{m}} \bm V_{lm}\bm x_{lm},
\end{equation}
where $\bm x_{lm}$ contains the data for each of the $S_{lm}$ streams of user $l$ in BS $m$, and $\bm V_{lm}$ is the $(M_K-N_f) \times S_{lm}$ complex precoding matrix of user $l$. For the sake of simplicity, $\bm P$ is assumed the same for all BSs with elements selected as $p_{ij}\ \in \mathbb{C}$ -- a truncated Fourier or Hadamard matrix being a good candidate in practice. The condition $N_f \geq 0$ in $\bm P$ means that the transmit signal at each BS occupies a reduced signal space of dimension $M_k-N_f$ and leaves the rest $N_f$ dimensions free. The maximum number of streams in the reduced space of a given cell is $S \defeq M_K-N_f$ and the DoF is equal to $\frac{M_K-N_f}{M_k}$. The received signal at the $l^{th}$ UE of the $m^{th}$ cell is given by
\begin{eqnarray}\label{eq2}
\bm y_{lm}&=&\underbrace{\bm H_{lm}\bm P\bm V_{lm}\bm x_{lm}}_\textrm{desired signal} + \underbrace{\bm H_{lm}\bm P \sum_{i\neq l}^{N_{m}}\bm V_{im}\bm x_{im}}_\textrm{intra-cell interference} \nonumber \\
&+& \underbrace{\sum_{i\neq m}^{B}\bm H_{li}\bm P\bm x_i}_\textrm{inter-cell interference}+\bm w_l\ ,
\end{eqnarray}
where $\bm H_{lm}$ is the direct channel matrix between the $m^{th}$ BS and the $ l^{th}$ UE, $\bm x_i = \sum_{n}^{N_i}\bm V_{in}\bm x_{in}$, and $\bm w_l$ is the $M_K$ complex circularly symmetric additive white Gaussian noise (AWGN) vector with zero mean and covariance matrix equals to $\sigma^2\bm I_{M_K}$. Each UE has perfect knowledge of at least the direct channel linking it to its main BS, and of its main interferers through some pilots as described e.g. in \cite{IRC}. In traditional transmission schemes, the model is nothing but \eqref{eq2} except that $\bm P$ is an identity matrix. This means that all dimensions are exploited at the expense of strong inter-cell interference at cell-edges, which results in poor performance. By using a truncated, rank-deficient $\bm P$, the BS can save some available dimensions to let the other BSs in the neighbors serve their users in an interference-free subspace. This technique allows each UE to have $N_f$ interference-free streams. The decoded signal at user $l$ of the $m^{th}$ BS is given by
\begin{eqnarray}\label{eq3}
\bm U_{lm}^{\dagger}\bm y_{lm}&=&\bm G_{lm}\bm V_{lm}\bm x_{lm} +  \bm G_{lm} \sum_{i\neq l}^{N_{m}}\bm V_{im}\bm x_{im} \nonumber\\
&+& \bm U_{lm}^{\dagger}\sum_{i\neq m}^{B}\bm H_{li}\bm P\bm x_i +\bm U_{lm}^{\dagger}\bm w_l\ ,
\end{eqnarray}
where $\bm G_{lm} = \bm U_{lm}^{\dagger}\bm H_{lm}\bm P$ is the direct equivalent channel between $l^{th}$ user and the $m^{th}$ BS.

\section{Interference Alignment in downlink}
\label{revscheme}
In \cite{SuhTse}, two IA schemes have been proposed, assuming only one stream per user. The first scheme is based on a Zero-Forcing (ZF) criterion. This means that each UE estimates the interfering channels and calculates the decoding matrix as their null space vector, i.e. $\bm U_{lm}$ collapses to a vector $\bm u_{lm}$ verifying: 
\begin{eqnarray}\label{eq4}
\bm u_{lm}^{\dagger}\sum_{i\neq m}^{N_{ri}}\bm H_{li}\bm P\bm x_i = 0,\ \ \textrm{subject to }\ ||u_{lm}|| = 1.
\end{eqnarray}
where $N_{ri}$ is the number of interfering BSs to be canceled. The basic case is obtained for $N_{ri} = 1$ which holds for the strongest interferer. Each user calculates the equivalent channel $\bm G_{lm}$ given in (\ref{eq3}) and feeds it back to its BS, which in turns apply a scheduler that selects a subset of users to serve. The selected users have their equivalent channels denoted $\bm{{\bar{c}}}_{lm}$ collected in a matrix as $\bm{\bar{C}}_m=\left[\bm{\bar{c}}_{1m},\cdots, \bm{\bar{c}}_{Sm}\right]$, and then a ZF beamforming scheme is applied to compute orthogonal transmission vectors avoiding intra-cell interference in \eqref{eq3}, 
\begin{equation}\label{eq5}
\left[\bm{\bar{v}}_{1m},\cdots, \bm{\bar{v}}_{Sm}\right] = \bm{\bar{C}}_{m}^{\dagger}\left(\bm{\bar{C}}_{m}\bm{\bar{C}}_{m}^{\dagger}\right)^{-1}.
\end{equation}
Each precoding vector is normalized to ensure a constant power: $\bm{\bar{v}}_{im}^0=\frac{\bm P\bm{\bar{v}}_{im}}{||\bm P\bm{\bar{v}}_{im}||}$.

The second scheme exploits a Minimum Mean Square Error (MMSE) criterion and takes into account the interference-plus-noise (IN) covariance in addition to the strongest interferer properties. The decoding vector is given by:
\begin{equation}\label{eq6}
\bm u_{lm} = \frac{\Phi_{lm}^{-1}\bm H_{lm}{\bm P}\bm v_{lm}^0 }{||\Phi_{lm}^{-1}\bm H_{lm}{\bm P}\bm v_{lm}^0||}.
\end{equation}
Assuming the knowledge of the strongest interferer channel, the IN covariance $\Phi_{lm}$ is
\begin{equation}\label{eq7}
\Phi_{lm} =  (\sigma^2+\textrm{INR}_\textrm{rem})\bm I_{M_K} + \frac{p}{S}\left(\bm H_{ln}{\bm P}\bm V_n\bm V_n^\dagger{\bm P}^{\dagger}\bm H_{ln}^{\dagger}\right),
\end{equation}
with $p$ the total transmit power, $\textrm{INR}_\textrm{rem}$ the remaining interference, and $\bm V_n = \left[\bm v_{1n},\cdots,\bm v_{Sn}\right]$ the precoding matrix of the $n^{th}$ BS. The initial vector $\bm v_{lm}^0$ can be selected so as to maximize the beamforming (BF) gain at the receiver, and chosen as the eigenvector associated to the maximum eigenvalue of $\bm H_{lm}^\dagger{\bm P}^\dagger\Phi_{lm}^{-1}\bm H_{lm}{\bm P}$. As can be seen from (\ref{eq6}) and (\ref{eq7}), one major difference comparing to the first scheme is that the interference-free subspace is no longer fixed and becomes dependent on the precoder choices of other BSs. However, in practice, we consider that in average, $\bm V_n\bm V_n^\dagger$ approaches an identity matrix because of the ZF BF step.
The strict reduction of the coding space with the matrix ${\bm P}$ can also be relaxed using a full square matrix with the last $N_f$ columns weighted by a factor $0<\kappa<1$. 
The motivation behind introducing the factor $\kappa$ is to be able to tweak the rank-deficiency of $\bm P$ and color the interference space \cite{SuhTse}. This approach gives more flexibility in the coding strategy and allows to adapt to situations where pure interference alignment is not necessary and may be outperformed by  interference management. The new mixing matrix ${\bm P}$ is given by 
\begin{equation}\label{eq8}
{\bm P} = \left[\bm p_1,\cdots,\bm p_{S},\kappa \bm p_{S+1},\cdots,\kappa \bm p_{M_K} \right],
\end{equation}
where the vectors $\bm p_i\ \in \mathbb{C}^{M_K}$ are mutually orthogonal.

\section{Generalized IA scheme}\label{genscheme}

The above IA schemes minimizes the dimensions lost in the signal space of each BS thanks to the controlled freed dimensions. While achieving a high SNR gain, it only assumes a stream per user, which severely restricts the dimensions of the problem. In many practical situations, we observed through simulations how this lack of diversity limits the system performance. Indeed, in each cell, some users are close to their main BSs and do not suffer from strong interference. In this case, it appears more efficient to let the user feeds back not only the interference-free signal direction but also additional directions presenting a sufficiently high overall SINR. By letting each user to feed back several directions, the scheduler has more flexibility to optimize the coding scheme. 

\subsection{Optimal coding directions}
Each UE first selects the optimal directions that maximize the received SINR, with  inter-cell interference only. The decoder is initialized as $\bm U_{lm} = \frac{\Phi_{lm}^{-1}\bm H_{lm}{\bm P}}{||\Phi_{lm}^{-1}\bm H_{lm}{\bm P}||}$. The equivalent channel is then given by
\begin{equation}\label{eq9}
\bm G_{lm} =  \bm P^\dagger\bm H_{lm}^\dagger\Phi_{lm}^{-1}\bm H_{lm}\bm P.
\end{equation}
The optimal receive directions are computed using the eigenvalue decomposition of $\bm G_{lm}$, with eigenvalues $\Lambda_{lm}$ and eigenvectors $\bm C_{lm}$ defined as 
\begin{equation}\label{eq10}
\bm C_{lm}^\dagger\Lambda_{lm}\bm C_{lm} = \bm G_{lm},
\end{equation}
where the elements of $\Lambda_{lm}$ are sorted in decreasing order, meaning that the first columns of $\bm C_{lm}$ provide the optimal receiving directions and may be used as decoding directions. In order to maximize the received SINR, the decoding matrix is chosen as $\bm U_{lm}^d = \frac{\Phi_{lm}^{-1}\bm H_{lm}\bm P\bm C_{lm}^d}{||\Phi_{lm}^{-1}\bm H_{lm}{\bm P}\bm C_{lm}^d||}$, and the precoding matrix as $\bm V_{lm}= \bm C_{lm}^d$, where $d$ is the number of selected streams for a given user. The resulting SINR of the $i^{th}$ stream can be easily found equal to $\lambda_{lm}^i$ the $i^{th}$ element of $\Lambda_{lm}$.

\subsection{Directions feedback}
Motivated by the fact that some users may not suffer a high interference level, we suggest that each user feeds the best $L \geq N_f$ directions calculated in (\ref{eq10}) back to its BS. In this way, UEs in the center of the cell preferably feed back decoding directions corresponding to strong eigenmodes of the direct channel, whereas cell-edge users feed back interference-minimizing directions. For the sake of consistency, the cell and user indexes are omitted in the upcoming equations. Each user thus feeds back a set of candidate decoding directions $\left\{(\bm c_1,\lambda_1),\cdots,(\bm c_{L},\lambda_{L})\right\},$
where $\bm c_i$ and $\lambda_i$ are the optimal coding direction and the SINR for the $i^{th}$ stream, respectively. It is worth noting that the $\lambda_i$ gain does not take into account the intra-cell interference, and indirectly is only an estimate of the SINR in the direction $\bm c_i$. When a BS serves only one user, the directions $\bm c_i$ are mutually orthogonal, and thereby there is no intra-cell interference. In the multi-user MIMO case, the directions for two different users, e.g. $\bm c_i$ and $\bm c'_j$ are not necessarily orthogonal, which adds an extra interference power of $\lambda_i||\bm c_i\bm c'_j||$. In order to cancel the intra-cell interference, we may call for a ZF-based precoding scheme as given in (\ref{eq5})\footnote{Other precoding types than the ZF can also be straightforwardly applied, see \cite{QS}. But this does not change the core of our work.}. However in doing so, the precoders are no longer perfectly aligned with the optimal receive directions, which introduce a potential rate loss in the network overall performance.

\subsection{Polynomial time sub-optimal scheduling}

\begin{algorithm}[b]
\caption{Sub-optimal heuristic algorithm}\label{heuristic}
\begin{algorithmic}[1]
\State $\mathcal{S}_u^0 \leftarrow \{\hat{s}_0\}$
\State $\mathcal{S}_c^0 \leftarrow \mathcal{S}-\{\hat{s}_0\}$
\State $Q_\text{opt} \leftarrow C(\{\hat{s}_0\})$\;
\While{$|\mathcal{S}_u^j| < S$ }
\For{$i \leftarrow 1:|\mathcal{S}_c^j|$}
    \State $Q \leftarrow C(\mathcal{S}_u^j \cup \{s_i\})$
    \If{$Q>Q_\text{opt}$}
       \State $Q_\text{opt} \leftarrow Q$
       \State $\hat{s}_j \leftarrow s_i$
    \EndIf
\EndFor
\State $j \leftarrow j+1$
\State $\mathcal{S}_u^{j}\leftarrow\mathcal{S}_u^{j-1} \cup \{\hat{s}_j\}$
\State $\mathcal{S}_c^{j}\leftarrow\mathcal{S}_c^{j-1} - \{\hat{s}_j\}$
\If{$\{\hat{s}_j\}\leftarrow\emptyset$}
	\State break 
\EndIf
\EndWhile
\end{algorithmic}
\end{algorithm}

The set of all candidates $\mathcal{S}=\left\{(\bm c_1,\lambda_1),\cdots,(\bm c_{N_c},\lambda_{N_c})\right\}$ is built by collecting all $N_c$ streams candidates from all users. Assuming a dense network scenario where $N_c$ is much greater than the number of available streams $S$, each BS has to select the best candidates in order to maximize a utility function defined as the sum-rate (with eventually fairness weights, see below). The optimal way is to apply an exhaustive search and to select the best subset $\mathcal{U}^*$ that results in the highest sum-rate. The optimization problem is defined as
\begin{equation}\label{eq12}
\mathcal{C(S)} = \underset{\mathcal{U} \in \mathcal{S}}{\max} \sum_{k\in \mathcal{U}} \omega_k\log_2 \left(1+\frac{p}{S}\lambda_k||\bm c_{k}\bm{\bar{v}}_{k}||\right),
\end{equation}
where $\mathcal{U}$ is a given candidates streams subset, and $\omega_k$ is a per-UE weight updated after each transmission in order to provide fairness among the different UEs. Since several directions are fed back per user, the dimensions of $\mathcal{S}$ increase exponentially, which in turns makes the optimal scheduling challenging as the network grows. The optimization problem in (\ref{eq12}) is in known as NP-hard, and its computational cost may be written as
\begin{equation}\label{eq13}
\text{cost} = \mathcal{O}\left(\binom {N_s}{S}\times\;\textrm{cost}_{zf}\right),%
\end{equation}
where cost$_{zf}$ is the complexity cost of a ZF precoding scheme. It is quite clear that for dense scenarios, the search among all candidate streams is not feasible. We rather propose a sub-optimal heuristic scheduling, that greatly decreases the complexity while preserving a near-optimal performance. For a given BS, the heuristic privileges the streams that provide higher rates and less inter-correlation.

Let us denote $\hat{s}_0$ the stream with the highest rate among all $N_c$ streams. At iteration $0$, we define the set of the chosen streams as $\mathcal{S}_u^0=\{\hat{s}_0\}$, the set of the remaining streams as $\mathcal{S}_c^0= \mathcal{S}-\{\hat{s}_0\}$, and the utility function as $C(\mathcal{S}_u^0)$ given in (\ref{eq12}). At the $i^{th}$ iteration, the utilities for $s \in \mathcal{S}_c^{i-1}$ are
\begin{equation}\label{eq14}
C(\mathcal{S}_u^i,s)=\sum_{k\in \mathcal{S}_u^{i-1}\cup\{s\}} \omega_k\log_2 \left(1+\frac{p}{S}\lambda_k||\bm c_{k}\bm{\bar{v}}_{k}||\right),
\end{equation}
and the stream $\hat{s}_i$ with maximum utility is selected. The sets of remaining and chosen streams are updated according to
\begin{eqnarray}\label{eq15}
\mathcal{S}_u^i&=&\mathcal{S}_u^{i-1} \cup \{\hat{s}_i\}, \nonumber \\
\mathcal{S}_c^i&=&\mathcal{S}_c^{i-1} - \{\hat{s}_i\}.
\end{eqnarray}
The algorithm stops when  no more gain can be achieved. In (\ref{eq14}), $\bm c_{k}$ is the fed back direction for a given stream, and $\bm{\bar{v}}_{k}$ results from the ZF precoding scheme applied on the vectors of $\mathcal{S}_u^{i-1}\cup\{s\}$. Looking at the computational cost of the heuristic algorithm given in Alg.\ref{heuristic}, we can readily see that the complexity is polynomial and the cost search process is upper-bounded by
\begin{equation}\label{eq16}
\text{cost}_\textrm{sub}=\mathcal{O}\left((N_s-\frac{S-1}{2})S\times\;\textrm{cost}_{zf}\right),
\end{equation}
since only two loops are required.

\section{Simulation Results}
\label{SR}
\begin{figure}[b]
\centering
\includegraphics[width=\columnwidth]{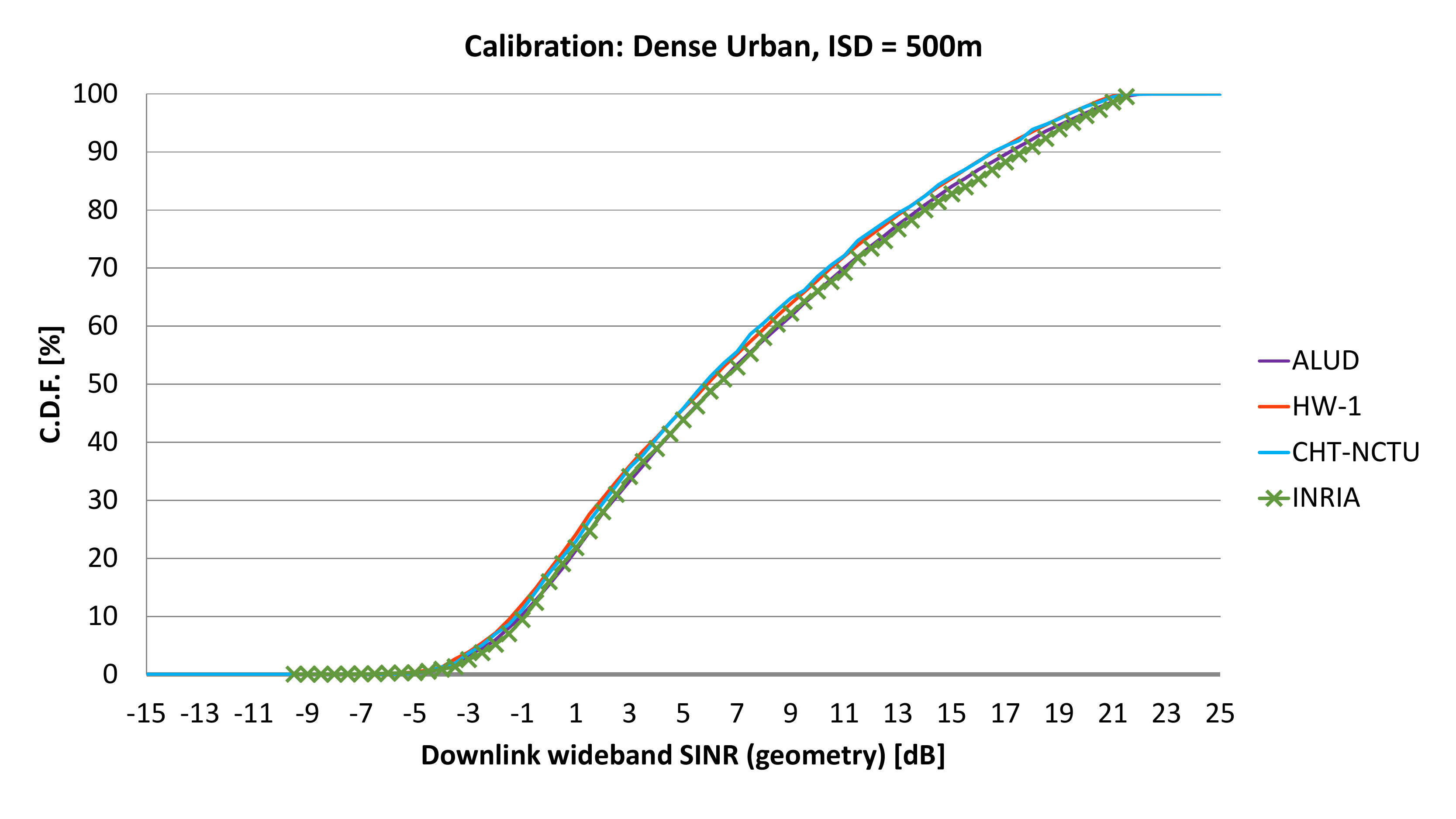}
\caption{Calibration results for our simulator (INRIA) with respect to other vendors and operators in the Greentouch consortium. This figure illustrates the distribution of the SINR experienced in average by the users, so-called the \emph{geometry}.}
\label{fig1}
\end{figure}

The performance of the proposed scheme has been evaluated through exhaustive simulations in different load scenarios. A system level simulator with $7$ cells is used, each cell being divided in $3$ sectors. For the exposed results, we only consider the center cell in which the UEs are uniformly distributed, although all cells support users and compute their precoders accordingly. We assume a proportional-fair scheduling by updating the scheduling weight of each user as
\begin{equation}
\omega_k^l=\frac{r_\textrm{min}}{\textrm{max}(r_\textrm{min},R_{k,\textrm{avg}}^l)},
\end{equation}
where $r_\textrm{min}$ is a threshold under which any rate is assumed null, and $R_{k,\textrm{avg}}^l$ is the average number of bits transmitted until the $l^{th}$ transmission. The spectral efficiency is compared to that of a classic OFDM scheme as a reference, where all interference are considered as noise. In this paper, $N_u=10$ users per cell in average are considered, and an overall coding space limited to $M_K=4$ dimensions with $K$ available sub-carriers and $M$ antennas at the BSs and the UEs. Our system level simulator assumes a regular placement of BSs and random users in full buffer. More information about the propagation model used here is given in \cite{GT} (cf GT doc2a). The performance are evaluated for both channel correlation levels: low and medium, with a correlation coefficient equal to $0$ and $0.3$, respectively. A simplified resource block structure is adopted to save computation time, the channel is abstracted as: $K$ sub-carriers with independent fading matrices, and each $M\times M$ fading matrix includes correlation. Each UE feeds back $M_K - N_f$ preferred receive directions. By assuming $M_K=4$ and $N_f=1$ free dimension at each BS, the BS chooses $S=3$ streams to be served using IA with scheduling based on Alg.\ref{heuristic}. The capacity is truncated to a maximum of $8$ bits per resource use. The Monte-Carlo simulations are run for $100$ independent scenarios. For each scenario, the geometry is fixed and the system is run for $100$ successive transmissions. As a reference, we provide the CDF of the received SINR in Fig.~\ref{fig1} for our simulator, calibrated to match system-level simulations of other partners in the Greentouch project. In the following, we assume that each UE has perfect estimate of the strongest interferer while it assumes others as noise.

\begin{figure}
\centering
\includegraphics[scale=0.24]{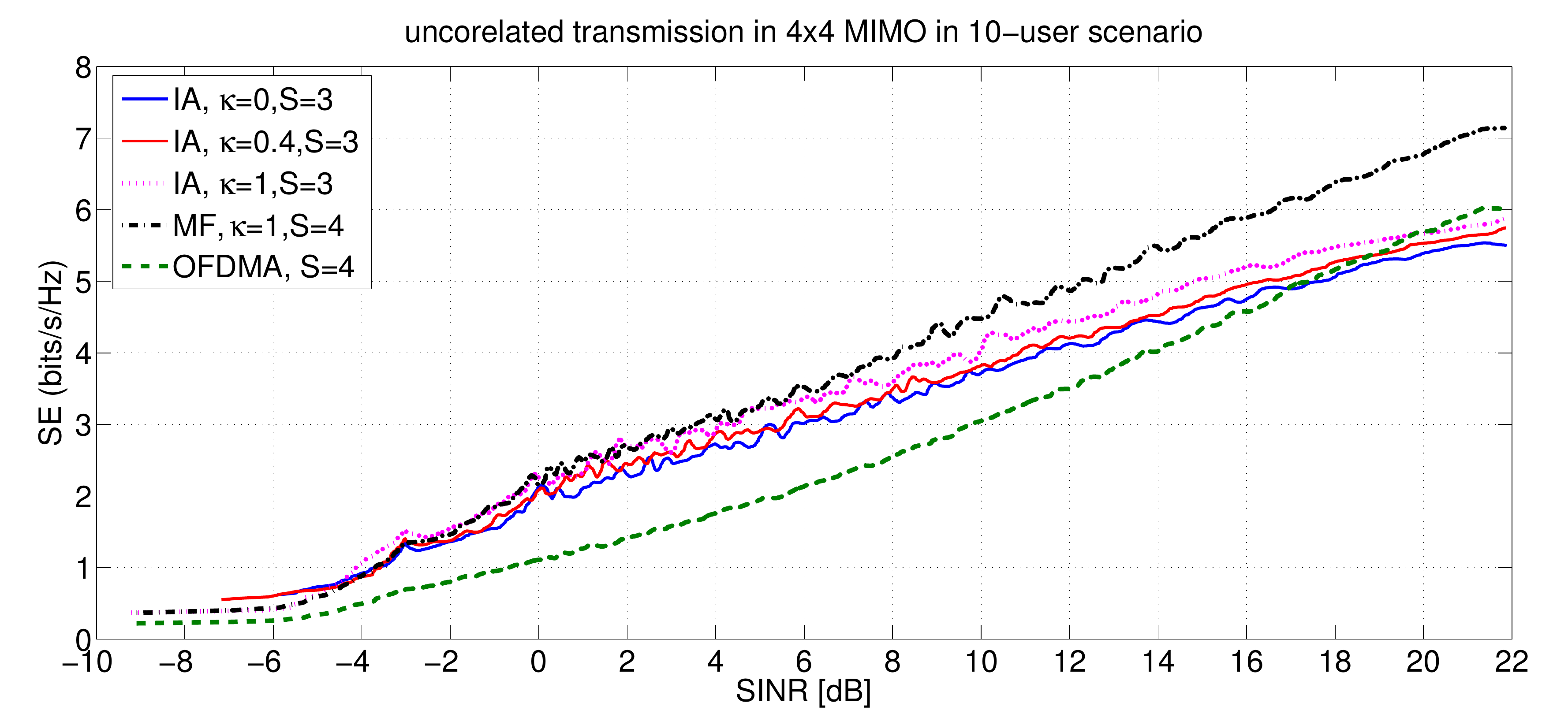}
\caption{Performance comparison of IA scheme for different $\kappa$ vs the matched filtering (MF) scheme and the basic MIMO-OFDM scheme in an uncorrelated scenario with $M=4$ , $K=1$.}
\label{fig2}
\end{figure}
In comparison to the reference scenario without IA in a $4\times 4$ uncorrelated MIMO scenario, one can observe from Fig.~\ref{fig2} a significant gain in the low to medium SINR region, e.g. the gain varies between $1 - 1.2$ bits/s/Hz at SINR $0$dB depending on the parameter $\kappa$. However, it decreases for high SINR because users are no longer suffering strong interference. Comparing to the MF based scheme where each BS exploits all dimensions leaving no free dimension for adjacent cells, a gain is only observed for very low SINR values i.e. SINR$<-4$dB. Beyond this value the MF scheme has an increasing gain specially for high SINR, where it also enjoys an almost constant gain over the OFDMA scheme. This means that due to channel diversity, a UE can frequently find a direction with low interference. The performance of the IA scheme for $\kappa=0$ and $\kappa=0.4$ are similar except that a slight gain is observed for SINR$>0$dB. Regarding the case with $\kappa=1$ and $S=3$, it can be seen that it outperforms the case where $\kappa=0$ and $\kappa=0.4$ for SINR$>2$dB. This is due to the additional dimension gained when $\kappa=1$. Similar observations have been made for the MIMO-OFDM configuration with $M=2$ and $K=2$. Except that in this case, the schemes with higher $\kappa$ always result in a SE gain. This means that the low-interference modes are basically created by the block diagonal channel, and are almost independent of the precoding design.

Contrarily to the uncorrelated configuration, in $2\times 2$ correlated channel with a factor of $0.3$ between the antennas and $K=2$ uncorrelated subcarriers, the performance are surprisingly different as shown in Fig. \ref{fig4}. For example, the IA schemes for different $\kappa$ greatly outperform the MF scheme for all SINR values, and result in significant gains compared to the OFDM scheme in the low-to-medium SINR region. This means that since all modes are correlated, an almost interference-free mode does no longer exist unless it is freed. And hence, the performance are degraded. However, comparing the IA scheme with different $\kappa$, it can be seen that $\kappa=1, S=3$ results in a rate-loss for SINR$<-3$dB, and yields a gain beyond $8$dB since more dimensions are available for precoding. 


\begin{figure}
\centering
\includegraphics[scale=0.24]{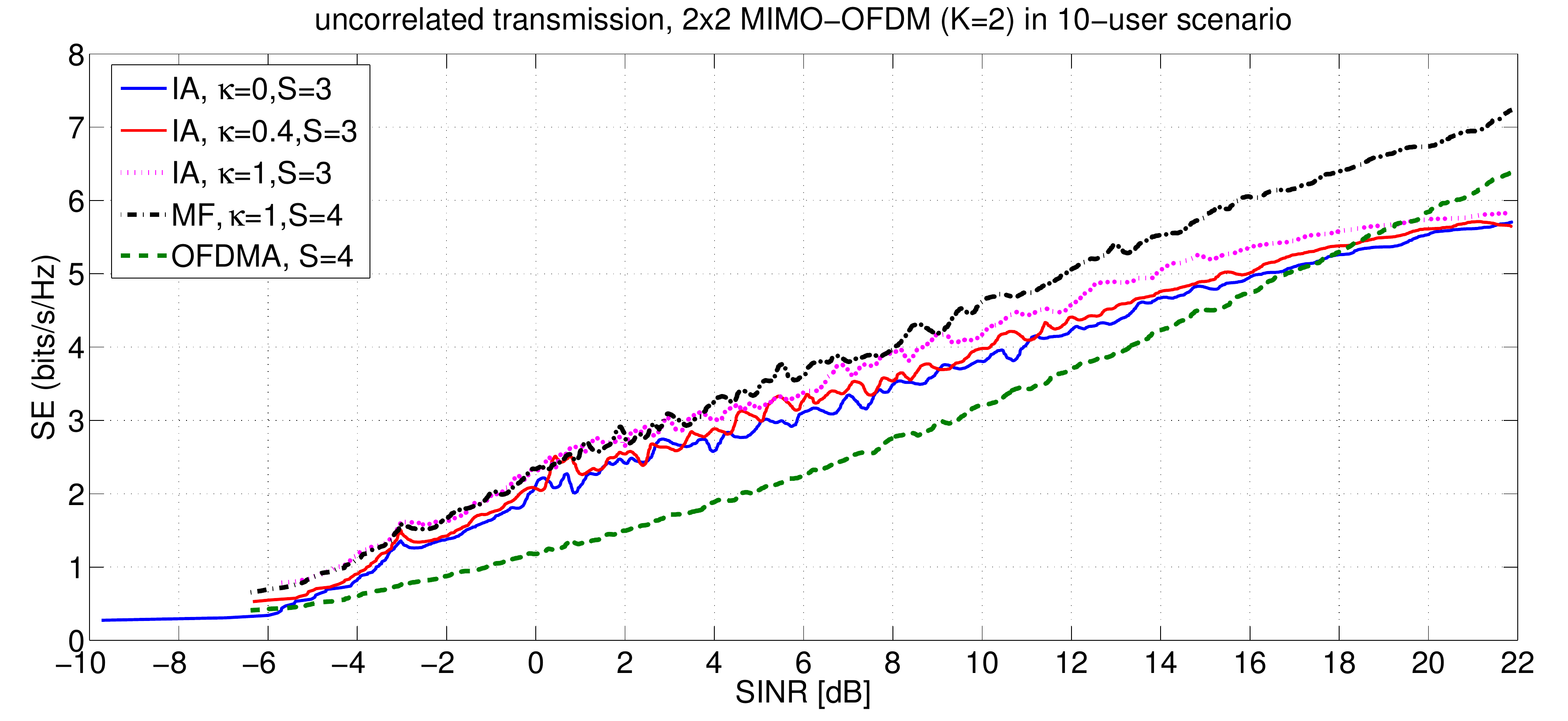}
\caption{Performance comparison for different $\kappa$ vs the MF and the basic MIMO-OFDM schemes in an uncorrelated scenario with $M=2$, $K=2$.}
\label{fig3}
\end{figure}

\begin{figure}
\centering
\includegraphics[scale=0.24]{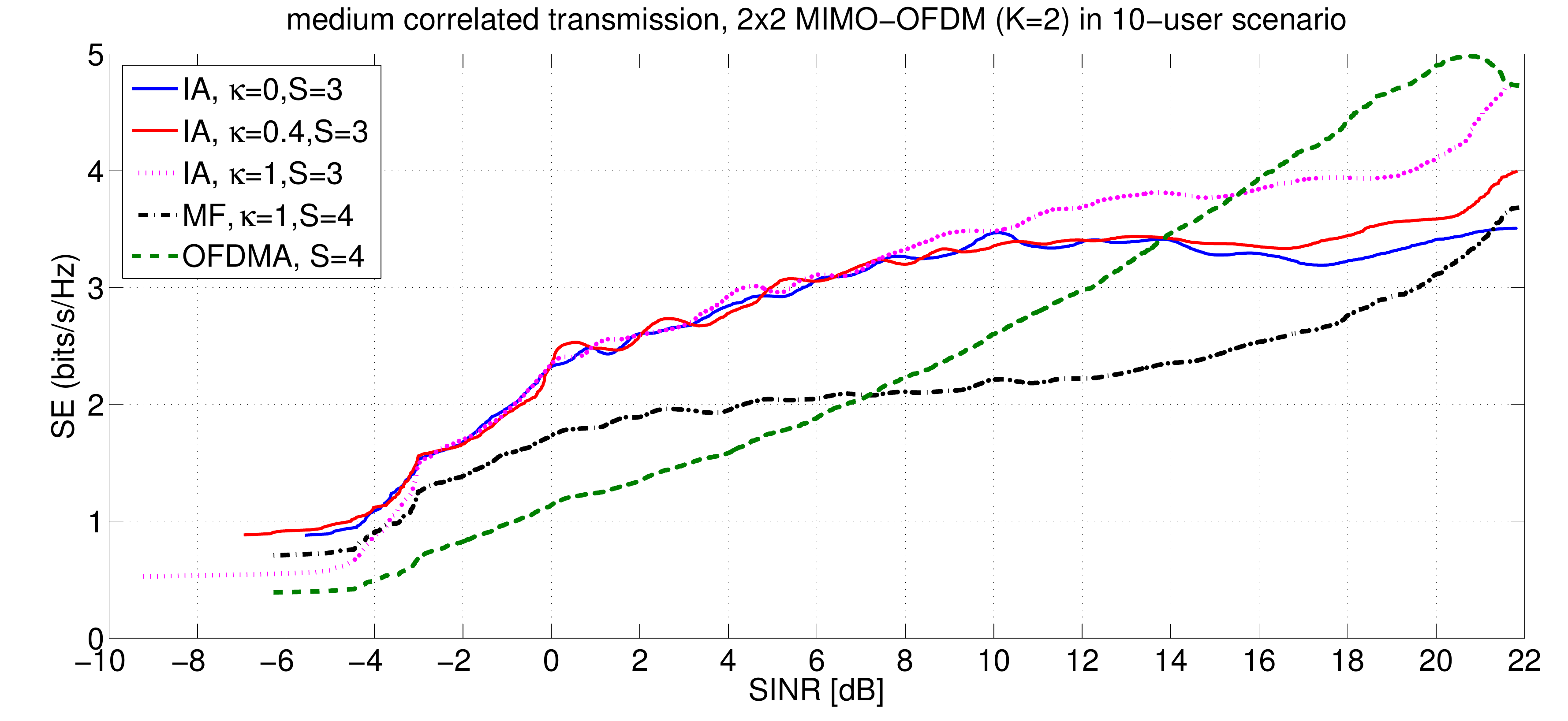}
\caption{Performance comparison for different $\kappa$ versus the matched filtering based scheme and the basic MIMO-OFDM reference scheme in a medium correlated scenario with $M=2$ and $K=2$.}
\label{fig4}
\end{figure}

\section{Conclusion}
This paper addressed the performance of IA transmission in a downlink cellular network. We have generalized the scheme proposed by \cite{SuhTse} and have came up with a near-optimal low-complexity scheduler based on heuristic optimization. We have also shown that unlike the results obtained in \cite{DA}, applying a joint scheduling-precoding based on IA transmission can yield significant gains for users suffering strong inter-cell interference. Also the IA scheme becomes valuable in correlated channels where the MF scheme yields poor performance. Future works should focus on avoiding the data-rate loss caused for users enjoying high SINR which could be obtained by tuning the fairness coefficients.

\end{document}